\documentclass[aps, pre, reprint, notitlepage, longbibliography, noeprint, superscriptaddress]{revtex4-2}
\usepackage{amsmath,amsfonts,amssymb,bm} 
\usepackage{xcolor,graphicx,bbold} 
\usepackage{microtype}
\usepackage[colorlinks=true,linkcolor=blue,citecolor=blue,urlcolor=blue]{hyperref}
\usepackage[all]{hypcap} 

\newcommand{\Dr}[0]{D_\mathrm{r}} 
 
\newcommand{\DcM}[0]{D_{\rm c}} 
\newcommand{\DaO}[0]{D_{\rm a}^{(0)}} 
\newcommand{\va}[0]{v_{\rm a}} 
\newcommand{\vaO}[0]{v_{\rm a}^{(0)}} 
\newcommand{\muc}[0]{\mu_{\rm c}} 
\newcommand{\ee}[0]{\mathrm{e}} 
\newcommand{\kB}[0]{k_{\rm B}} 
\newcommand{\rvec}[0]{{\bm r}} 
\newcommand{\nvec}[0]{{\bm n}} 

\begin{document}
\title{Enhanced diffusivity in microscopically reversible active matter}

\author{Artem Ryabov} 
\email[]{artem.ryabov@mff.cuni.cz}
\affiliation{Charles University, Faculty of Mathematics and Physics, Department of Macromolecular Physics, V Hole{\v s}ovi{\v c}k{\' a}ch~2, 180~00~Praha~8, Czech Republic} 
\affiliation{Departamento de F\'{\i}sica, Faculdade de Ci\^{e}ncias, Universidade de Lisboa, 
1749-016 Lisboa, Portugal}
\affiliation{Centro de F\'{i}sica Te\'{o}rica e Computacional, 
  Faculdade de Ci\^{e}ncias, Universidade de Lisboa, 1749-016 Lisboa, Portugal}

\author{Mykola Tasinkevych}
\email[]{mykola.tasinkevych@ntu.ac.uk}
\affiliation{SOFT Group, School of Science and Technology, Nottingham Trent University, Clifton Lane, Nottingham NG11 8NS, UK}
\affiliation{Departamento de F\'{\i}sica, Faculdade de Ci\^{e}ncias, Universidade de Lisboa, 
1749-016 Lisboa, Portugal}
\affiliation{Centro de F\'{i}sica Te\'{o}rica e Computacional, 
Faculdade de Ci\^{e}ncias, Universidade de Lisboa, 1749-016 Lisboa, Portugal}

\date{March 03, 2022} 

\begin{abstract}
Physics of self-propelled objects at the nanoscale is a rapidly developing research field where recent experiments focused on motion of individual catalytic enzymes. Contrary to the experimental advancements, theoretical understanding of possible self-propulsion mechanisms at these scales is limited. A particularly puzzling question concerns origins of reportedly high diffusivities of the individual enzymes. Here we start with the fundamental principle of microscopic reversibility (MR) of chemical reactions powering the self-propulsion and demonstrate that MR can lead to increase of the particle mobility and of short- and long-time diffusion coefficients as compared to dynamics where MR is neglected. Furthermore, the derived diffusion coefficients are enhanced due to the action of an external force. These results can shed new light on interpretations of the measured diffusivities and help to test the relevance of MR for the active motion of individual nanoswimmers.
\end{abstract}

\maketitle 

\section{Introduction}

Active particles convert free energy of the host medium directly into mechanical energy of translation and rotation \cite{Bechinger2016}, which provides an opportunity to use them as microrobots \cite{Soto:2021} in advanced applications such as transport of microcargoes in lab-on-chip devices \cite{Baraban2012}, drug delivery \cite{Patra2013,Xu2020}, biosensing \cite{Wu:2010}, environmental remediation \cite{Soler2013,Soler2014}, or assembly of microstructures \cite{Sanchez2015}. From a basic science perspective, the active self-propelled motion leads to novel types of emergent collective behavior not accessible in passive systems. This includes so called living crystals \cite{Palacci2013} and motility-induced phase separation of particles with repulsive interactions only \cite{Fily2012, Redner2013, Levis2014}. With all the fascinating properties, active matter has become an exciting research subject for non-equilibrium statistical physics. 

Basic mechanisms underlying self-propulsion at length scales of several micrometers and above are understood rather well. At these scales, the active motion is typically powered by self-phoresis \cite{anderson89, golestanian05, popescu10, Illien17, Kroy:2016, Ma:2017}, i.e., by means of tangential surface flows due to self-induced gradients of the product and/or reactant molecules \cite{Howse2007, golestanian07, popescu10} or the electric potential \cite{Moran11,ebbens14,brown14,Brown17,Moran17}. Their theoretical description can be carried out at the hydrodynamic level \cite{ibrahim_golestanian_liverpool_2017, Illien:2017b, Kroy2018, Kanso2019} disregarding fluctuations in the number of chemical reactions, rendering thus a non-stochastic self-propulsion velocity.  

Contrary, the fluctuations are inherently present at nanometer scales where recent experimental studies advocated that enzymes can exhibit self-propulsion when catalyzing their respective  reactions \cite{Muddana:2010, Sengupta:2013, Sengupta:2014, Jee:2018a, Ah-Young/etal:2018,Jee:2020,Yuan:2021}. The idea of self-propulsion was invoked to explain reportedly high diffusion coefficients of a wide range of enzymes. However, models used to interpret measured data either neglect fluctuations of the active velocity or do not consider the effect of microscopic reversibility of chemical reactions and underestimate the reported values \cite{FENG2019, Feng/Gilson:2019, Feng:2020}. Despite growing number of experimental and theoretical studies,  the physical origins behind experimental observations on active enzymes \cite{Wang/etal:2020, Gunther/etal:2020, Wang/etal:2020:2} and their possible self-propulsion mechanism \cite{Feng/Gilson:2019, Feng:2020} are still under active debates. 

In this article, assuming that such a mechanism does exist, we focus on the fundamental principle of microscopic reversibility (MR)~\cite{Onsager:1931a, Astumian:2015} applied to chemical reactions powering the self-propulsion. We show that MR has a pronounced impact on dynamics of the nanoswimmer and leads to enhancements of its mobility and diffusivity compared to a passive particle and to models where MR is neglected. Moreover, both quantities become sensitive to the externally applied forces. We explain microscopic origins of these effects on a minimal thermodynamically consistent Markovian model of self-propulsion velocity. Our main findings are formulated as specific proposals for experimental tests aimed to verify the effects of MR. At the same time, they suggest a way for controlling particle mobility and diffusivity and provide insights into further possible reasons for occurrence of the enhanced diffusivities at the nanoscale.

\section{Microscopically reversible active propulsion} 
\begin{figure}[t]
\centering
\includegraphics[width=0.6\columnwidth]{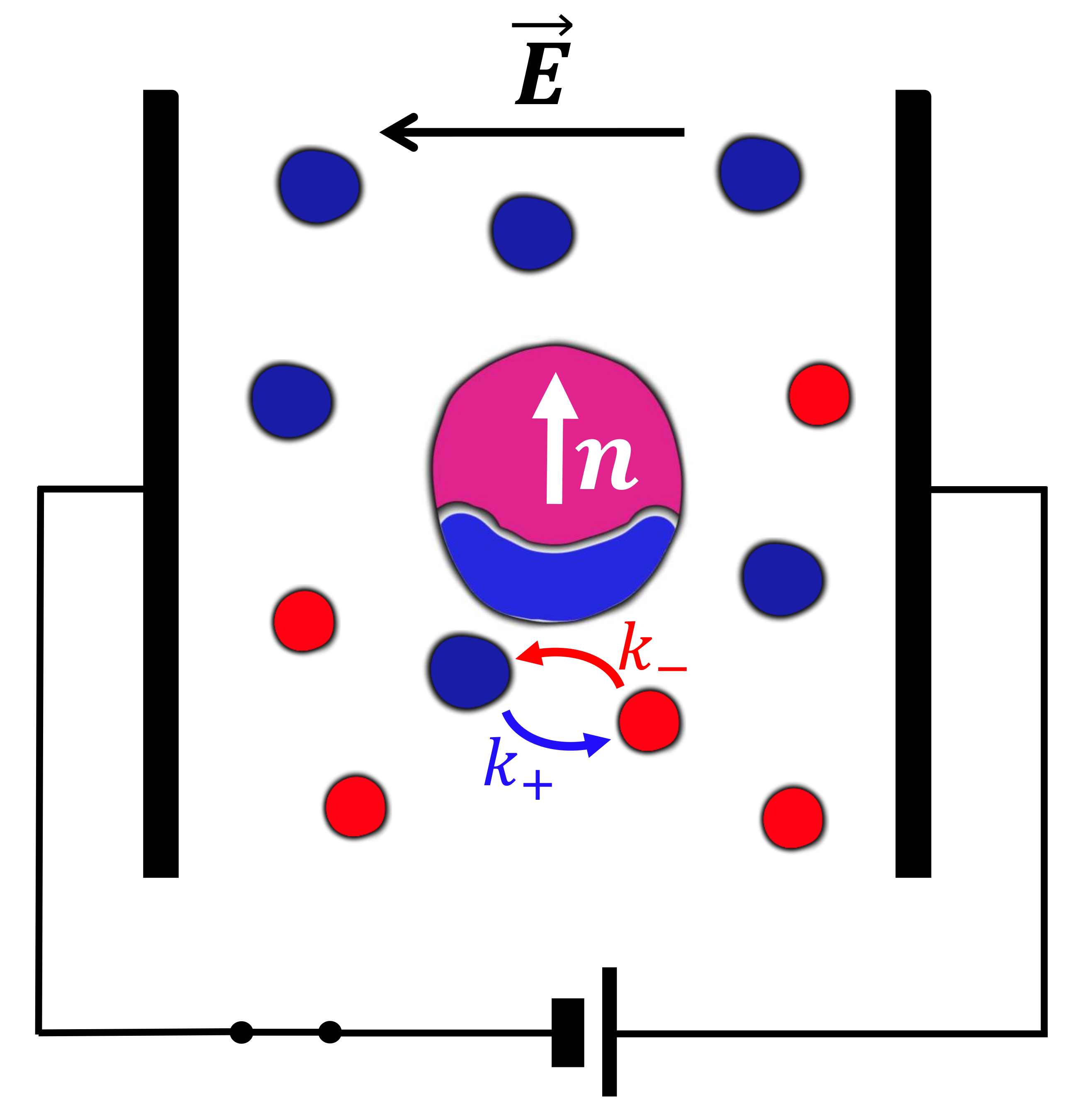}
\caption{Schematic illustration of a catalytic nanoswimmer. The catalytic region of the swimmer, depicted in blue, promotes the forward reaction step of the decomposition of a substrate molecule (small blue circle) into a product (small red circle) molecule. The rate of the forward steps is $k_+$ and each step causes particle displacement by $\delta r>0$ along the particle orientation $\nvec$. The reversed reaction steps with the rate $k_-$  transform the product molecules back into the substrate driving particle translation by $\delta r$ in the opposite $-\nvec$ direction. As an example, it is illustrated here that the nanoswimmer is electrically charged. The source of an external constant force is schematically represented by an electric field generator. Then applying an external electric field will result in an increase of the diffusion coefficient of the nanoswimmer as compared to the case of zero field. The enhancement of the diffusion coefficient $\propto F^2$ at leading order in $F$, where $F$ is a magnitude of the force exerted upon the nanoswimmer. This result holds in general, not only for electric forces.}
\label{fig:schme}
\end{figure}

Consider a nanoswimmer whose active propulsion is based on a chemical reaction that dissipates the free energy $\Delta G_{\rm r}$. Each such reaction causes a displacement $\delta r$ along the particle orientation $\nvec$ from the initial particle position $\rvec$ to $\rvec + \nvec \delta r$. If the particle moves in an external potential $V(\rvec )$ (e.g., in a potential in optical tweezers, the one of gravitational, or electrostatic forces; figure \ref{fig:schme} schematically shows such an active particle in an external electric field), in each such displacement the work $\delta W = V(\rvec + \nvec \delta r )-V(\rvec )$ is done against the external force ${\bm F} = -\nabla V$.  
Microscopic reversibility of the chemical reaction~\cite{Onsager:1931a} guarantees that to each forward reaction leading to the displacement $+\delta r$, there exists a reversed one associated with $-\delta r$, see schematics in Fig.~\ref{fig:schme}. Moreover, rates of the forward ($k_+$) and reversed ($k_-$) reactions obey the local detailed balance condition  
\begin{equation}
\frac{k_+}{k_-} = 
\ee^{( \Delta G_{\rm r} - \delta W) / \kB T },
\label{eq:detailed_balance}
\end{equation} 
where $\kB$ is the Boltzmann constant and $T$ is the temperature of ambient environment.  

As a result of this active process and the passive diffusion, the particle's center of mass position $\rvec$ evolves in time according to the Langevin equation   
\begin{align} 
\label{eq:Langevin_exact}
\begin{split} 
\frac{d \rvec }{dt} = u_{\rm a}(t) \nvec(t) + 
 \mu {\bm F}(\rvec) + \sqrt{2 D }\, {\bm \xi}(t) , 
\end{split} 
\end{align} 
where the magnitude of active velocity, $u_{\rm a}(t)$, is a (time derivative of) Markov jump process with detailed balanced rates $k_\pm$. It describes active jumps by $\pm\delta r$ along the particle orientation $\nvec(t)$, which changes over time due to a rotational diffusion with the diffusion coefficient $\Dr$. 
For the sake of simplicity we consider two-dimensional dynamics and a spherically symmetric particle. 
Then ${\bm n}(t)=( \cos\phi(t) ,\sin\phi(t) )$, and increments of the angle $\phi(t)$ arise from integration of the delta-correlated zero-mean Gaussian white noise $\xi_{\rm r}(t)$:
$\phi(t) = \phi(0) + \sqrt{2 \Dr }\int_0^t dt' \xi_{\rm r}(t')$. 

The last two terms on the right-hand side of~\eqref{eq:Langevin_exact} describe the overdamped passive Brownian motion in the force field ${\bm F}(\rvec)$ with the translational diffusion coefficient $D$ and mobility $\mu$ satisfy related by the fluctuation-dissipation theorem $D=\mu \kB T$. 
Thermal noise components, ${\bm \xi}(t)=(\xi_x(t),\xi_y(t))$, are zero-mean delta-correlated Gaussian processes: 
$\langle \xi_i(t) \rangle=0$, $\langle \xi_i(t_1)\xi_j(t_2) \rangle=\delta_{ij} \delta(t_1-t_2)$, $i,j=x,y$. 

Notice that models of active particles with the thermodynamically consistent propulsion, i.e., obeying the detailed balance condition~\eqref{eq:detailed_balance}, have appeared only recently \cite{Pietzonka/Seifert:2018, Speck:2018}. They were invoked in the discussion of the entropy production \cite{Pietzonka/Seifert:2018}, establishing a stochastic-thermodynamic description of the active motion \cite{Speck:2018}, derivation of the phoretic velocity \cite{Speck:2019}, discussion of motility-induced phase separations \cite{Fisher/etal:2019}, and of active heat engines \cite{Pietzonka/etal:2019}. Further thermodynamically consistent models, where chemical kinetics is also modeled explicitly, were discussed in the linear-response regime ($\Delta G_{\rm r} / \kB T \ll 1$) \cite{Gaspard/Kapral:2017JCP, gaspar:2018, huang:2018}.

\section{Enhanced mobility and force-dependent diffusivity} 
Even for a constant force $\bm F(\rvec)=(F,0)$ acting upon the particle center of mass, 
the work $\delta W$ in the detailed balance condition~\eqref{eq:detailed_balance} depends on time, $\delta W = \delta W(t)$,  
\begin{equation}
\label{eq:dW}
\delta W(t) = - F_{n}(t)\delta r= - F \delta r \cos \phi(t), 
\end{equation} 
where  $F_n(t)=\nvec(t)\cdot \bm F$ denotes the force component parallel with the particle orientation. Hence in the constant force field, 
the both rates $k_\pm$ in~\eqref{eq:detailed_balance} can be time-dependent, $k_\pm=k_\pm(t)$ as a consequence of MR.

To identify measurable effects that MR has on the particle dynamics, let us focus on the first and the second moments of the particle position $\rvec(t) = (x(t),y(t))$. 
We derive them after solving a master equation corresponding to \eqref{eq:Langevin_exact}. Details of the derivation are presented in the Electronic Supplementary Information (ESI). As a first and general result, we obtain 
$\langle \rvec(t) \rangle$ and $\langle \rvec^2(t) \rangle$ for arbitrary time-dependent rates $k_\pm(t)$. 

In particular, assuming initially the particle to reside at the origin [$x(0)=0$, $y(0)=0$], and its orientation to be random [$\phi(0) \in [0,2\pi)$], for its mean coordinates we get    
\begin{align} 
\label{eq:x_mean_CF}
& \langle x(t) \rangle = \mu F t + 
\int_0^t dt' \langle \va (t') \cos \phi(t')  \rangle ,\\
\label{eq:y_mean_CF} 
& \langle y(t) \rangle =  0. 
\end{align} 
The average under the integral in~\eqref{eq:x_mean_CF} is taken with respect to the distribution of process $\phi(t)$, with $\va(t)$ being the mean active velocity conditioned on the fixed particle orientation. It is given by 
\begin{equation}
\va(t) = \delta r [ k_+(t) - k_-(t)]. 
\end{equation}

After introducing dispersion of the active displacement for a fixed orientation $D_{\rm a} (t)$,
\begin{equation}
D_{\rm a} (t) = \frac{(\delta r)^2}{2}[ k_+(t) + k_-(t)],
\end{equation}
we obtain the variance of the particle position:
\begin{align}
\begin{split}
\label{eq:var_r} 
& \langle \rvec^2(t) \rangle - \langle \rvec(t) \rangle^2 = 4Dt + 2 \int_0^t dt' \langle D_{\rm a} (t') \rangle 
+  \\ 
 &  + \int_0^t dt_1 \int_0^t dt_2 \biggl \{ \langle \va (t_1) \va(t_2) \cos[ \phi (t_2)-\phi (t_1)]  \rangle \\
& \qquad \qquad - \langle \va (t_2) \cos \phi (t_2) \rangle^2 \biggr \}. 
\end{split}
\end{align}
Also in this result, the averages on the right-hand side are over realizations of the process $\phi(t)$. 

For a further analysis of general results~\eqref{eq:x_mean_CF} and~\eqref{eq:var_r}, we need to specify dependencies of rates $k_\pm(t)$ on the work function $\delta W(t)$ from \eqref{eq:dW}. Usually, finding an exact form of $k_\pm(t)$ is a model-specific task~\cite{Hill:1974, Astumian:2016} particularly challenging when starting from a microscopic dynamics \cite{Nitzan:2006, Gaspard/Kapral:2019}. 
Instead of proceeding along this direction and thus discussing a specific model, 
let us assume the active propulsion to be a thermally activated process. 
In such a case, detailed balanced activated rates are frequently chosen in the exponential form 
\begin{equation} 
\label{eq:kpm_W}
k_\pm(t) =k_\pm^{(0)} \exp\!\left[ \mp \frac{\delta W(t)}{2\kB T} \right], 
\end{equation} 
where constant rates $k_\pm^{(0)}$, satisfying $k_+^{(0)}/k_-^{(0)}=\exp[\Delta G_{\rm r}/\kB T]$, describe the active propulsion in absence of the external force. The factor 1/2 in the exponent in~\eqref{eq:kpm_W} assures that these rates are detailed balanced according to Eq.~\eqref{eq:detailed_balance}. 

For the sake of clarity, we will discuss properties of Eqs.~\eqref{eq:x_mean_CF} and~\eqref{eq:var_r}  in the weak-force limit: $F \delta r/2 \kB T < 1$, 
see formulas~(\ref{eq:xmean_exact})-(\ref{eq:MSD_smalltime_exact}), and~\eqref{eq:D_full} below. 
If $\delta r$ is in nanometers and $\kB T$ corresponds to the room temperature, this inequality is satisfied by $F$ of the order of piconewtons or weaker. 
In case of stronger forces, exact formulas valid for arbitrary $F$ given in Sec.~I of ESI should be used. 
Physical meanings of the exact formulas and their dependence on parameters are qualitatively similar to those of approximate ones provided below. The main difference is that exact results contain (series of) modified Bessel functions of the first kind of the argument $(F \delta r/2 \kB T)$ instead of linear and quadratic functions displayed in Eqs.~(\ref{eq:xmean_exact})-(\ref{eq:MSD_smalltime_exact}), and in~\eqref{eq:D_full}. 

Notably, a specific form of $k_\pm^{(0)}$ is not needed for the evaluation of averages on the right-hand side of Eqs.~\eqref{eq:x_mean_CF} and~\eqref{eq:var_r}. All results are expressed in terms of the mean active velocity $\vaO$ and the active diffusivity $\DaO$ of a force-free dynamics defined as 
\begin{align}
\label{eq:vaO}  
& \vaO = \delta r (k_+^{(0)}-k_-^{(0)} ),\\ 
\label{eq:DaO}
& \DaO = \frac{(\delta r)^2}{2} (k_+^{(0)}+k_-^{(0)} ), 
\end{align} 
respectively. 

If MR is neglected, \eqref{eq:x_mean_CF} gives us response of $\langle x(t) \rangle$ to the applied force identical to that of a passive Brownian particle: $\langle x(t) \rangle = \mu F t$. 
Indeed, neglecting MR means disregarding the mechanochemical coupling required by MR in~\eqref{eq:detailed_balance}. In such case, the rates $k_\pm(t)$ do not depend on the mechanical work $\delta W(t)$, $k_\pm(t)=k_\pm^{(0)}$, hence $\va(t')=\va^{(0)}$ is constant and the correlation function $\langle \va(t') \cos \phi(t')  \rangle $ in~\eqref{eq:x_mean_CF} vanishes because $\langle \cos \phi(t')  \rangle=0 $.  

For the microscopically reversible active propulsion, \eqref{eq:x_mean_CF} gives us    
\begin{equation} 
\label{eq:xmean_exact}
\langle x(t) \rangle = \left( \mu + \frac{1}{2} \frac{\DaO}{\kB T} \right ) F t , 
\end{equation} 
where we can identify $\DaO/\kB T$ as the active mobility $\mu_{\rm a}^{(0)}= \DaO/\kB T$  in accordance with the fluctuation-dissipation theorem. 
This extra term stems from dependence of $v_{\rm a}(t)$ on the particle orientation through $\delta W(t)$. 
The velocity $v_{\rm a}(t)$ as well as $\cos \phi(t)$ are largest when the particle is oriented in parallel with $\bm F$, i.e., when $\cos \phi(t) =1$. 
Contrary, there is no effect of $\bm F$ on $v_{\rm a}(t)$ for the perpendicular orientation [$\cos \phi(t) =0$]. Averaging over all orientations leads to the factor 1/2 in~\eqref{eq:xmean_exact}. 

Let us now describe fluctuations of $\rvec(t)$ around its mean value. They arise from three qualitatively different sources as quantified by the three terms in \eqref{eq:var_r}. 
The first term, $4Dt$, corresponds to the passive Brownian motion. The second, $2 \int_0^t dt' \langle D_{\rm a} (t') \rangle$, originates from fluctuations in number of chemical reactions powering the active propulsion. In fact, $\langle D_{\rm a} (t) \rangle$ does not depend on time and we have  
\begin{equation}
\label{eq:int_Da}
\int_0^t dt' \langle D_{\rm a} (t') \rangle = 
\DaO \left[1 + \left( \frac{F\delta r}{4\kB T} \right)^2 \right] t. 
\end{equation}
The third term on the right-hand side of~\eqref{eq:var_r} accounts for the amount of noise added to statistics of $\rvec(t)$ due to the rotational diffusion of particle's active velocity. 

In experiments, the fluctuations of $\rvec(t)$ are most widely discussed within the short-time and the long-time limits \cite{Jee:2018a,Ah-Young/etal:2018,Jee:2020}. In both regimes, the variance $\langle \rvec^2(t) \rangle - \langle \rvec(t) \rangle^2 $ grows linearly with $t$. Diffusivities, giving slopes of the linear $t$-dependencies within these regimes, differ by the third, ``rotational-diffusion'', term in~\eqref{eq:var_r}. 

In the short-time limit, the result~\eqref{eq:var_r} yields  
\begin{equation} 
\label{eq:MSD_smalltime_exact}
\langle \rvec^2(t) \rangle - \langle \rvec(t) \rangle^2 \approx 
4 \left\{ D + \frac{\DaO}{2}\left[1 + \left( \frac{F\delta r}{4\kB T} \right)^2 \right] \right\}  t,
\end{equation}  
where we have neglected higher-order powers of $t$, which are all in the form $(\Dr t)^m$, $m=2,3,\ldots$, and correspond to power series expansion of the third term in~\eqref{eq:var_r}. Such limit means $t\ll \Dr ^{-1}$, i.e., the rotational diffusion is negligible in this regime. Therefore, the enhancement of short-time diffusivity in~\eqref{eq:MSD_smalltime_exact} (as compared to the passive Brownian motion) is entirely due to~\eqref{eq:int_Da}. 

For long observation times ($t\gg \Dr^{-1}$), \eqref{eq:var_r} predicts the effective diffusion coefficient of the particle,  
\begin{equation}
\label{eq:D_definition}
\mathcal D = \lim_{t \to \infty} 
\frac{\langle \rvec^2(t) \rangle - \langle \rvec(t) \rangle^2 }{ 4 t}, 
\end{equation}
to be given by 
\begin{align}
\label{eq:D_full} 
\mathcal{D} &= D + \frac{\DaO}{2} \left[1+ \left( \frac{F\delta r}{4\kB T} \right)^2 \right] \\ \nonumber
& \quad  
+ \frac{(\vaO)^2}{2\Dr} \left[1 + \frac{1}{8} \left( \frac{F\delta r}{\kB T} \right)^2 \right] 
 +\frac{1}{32\Dr}  \left( \frac{\DaO F }{\kB T} \right)^2 .
\end{align} 
Here, the first line is equal to the short-time diffusivity, cf.~\eqref{eq:MSD_smalltime_exact}. The $\Dr$-dependent terms on the second line originate from the third term on the right-hand side of~\eqref{eq:var_r}. 

Remarkably, in the both regimes~\eqref{eq:MSD_smalltime_exact} and~\eqref{eq:D_full}, the particle diffusion constant increases with the amplitude of the external force $F$. The force-dependent terms responsible for the enhancement would be absent in a phenomenological model neglecting MR and assuming $\delta W$-independent rates 
$k_\pm(t)=k_\pm^{(0)}$ for the active propulsion. 

\section{Active propulsion in the macroscopic limit} 
In the Langevin equation~\eqref{eq:Langevin_exact}, $u_{\rm a}(t)$ is a random shot noise-like process reflecting the time-discrete nature of chemical reactions. 
If the active motion is observed on macroscopic time and length scales, 
$u_{\rm a}(t)$ becomes 
\begin{equation}
\label{eq:ua_macro}
u_{\rm a}(t) \approx u + \muc F_{n}(t) + \sqrt{2 \DcM}\, \xi_n(t) ,
\end{equation} 
which depends on the constant mean active velocity $u$, given as the free energy of reaction per elementary displacement $ (u/\muc) \delta r = \Delta G_{\rm r}$, on the associated mobility $\muc$ on the external force projected onto particle's orientation, $F_n(t)$, introduced in~\eqref{eq:dW}, and on the zero-mean Gaussian white noise $\xi_n(t)$ with the intensity controlled by $\DcM=\muc \kB T$. The white noise describes fluctuations in the number of chemical reactions as observed at the macroscale. Notably, if $\bm F =0$ ($F_{n}=0$), the Langevin equation~\eqref{eq:Langevin_exact} with the coarsegrained active velocity~\eqref{eq:ua_macro} can be formally mapped onto that of an asymmetric active Brownian particle \cite{Kurzthaler/etal:2016}.

A derivation of the macroscopic approximation~\eqref{eq:ua_macro} starting from the microscopic dynamics is presented in ESI. It is formally performed by taking the  $\delta r\to 0$ limit while both $\vaO$ and $\DaO$, given in Eqs.~\eqref{eq:vaO} and \eqref{eq:DaO}, remain finite and nonzero. This is achieved by imposing the usual diffusive coupling of microscopic time and length scales, i.e., ``(position)$^2$ $\sim$ time'' or $k_\pm \sim (1/\delta r)^2$. The macroscopic quantity $\DcM$ is then obtained from the microscopic model as the $\delta r\to 0$ limit of  $\DaO$.  

In practice, if rate constants describing the active propulsion are not known, the diffusion coefficient $\DcM$ (or $\muc = \DcM/\kB T$) should be treated as a phenomenological macroscopic parameter. Its value can be determined experimentally assuming the value of passive mobility $\mu$ of the particle is known. To this end, one can measure the mean position along the force direction and infer $\muc$ from the result~\eqref{eq:x_mean_macro}. Or/and one can determine the short- and long-time diffusivities and compare their values with Eqs.~\eqref{eq:MSD_smalltime_macro} and~\eqref{eq:D_macro}, respectively.

Macroscopic expressions for the moments of $\rvec(t)$ can be derived by solving the Langevin equation~\eqref{eq:Langevin_exact} with the macroscopic active velocity~\eqref{eq:ua_macro}; mathematical details are included in ESI. 
As a result, the macroscopic behavior of $\langle x(t)\rangle$ is similar to its microscopic counterpart~\eqref{eq:xmean_exact}, 
\begin{align}
\label{eq:x_mean_macro}
\langle x(t)\rangle & = \left(\mu + \frac{\muc}{2}\right)Ft,\\ 
\langle y(t)\rangle & =0.
\end{align} 
The particle mobility parallel to the force is thus enhanced as compared to the passive mobility $\mu$. 

For the short-time fluctuations of $\rvec(t)$, we get   
\begin{equation}
\label{eq:MSD_smalltime_macro} 
\langle \rvec^2(t) \rangle - \langle \rvec(t) \rangle^2  
\approx 4 \left( D + \frac{\DcM}{2} \right) t  , \quad \Dr t \ll 1,
\end{equation} 
and for the long-time diffusion constant~\eqref{eq:D_definition}, which we denote as $\mathcal{D}_{\rm m}$ in the macroscopic limit, we obtain 
\begin{equation}
\label{eq:D_macro} 
\mathcal{D}_{\rm m}=
D + \frac{\DcM}{2} + \frac{u^2}{2\Dr} + \frac{\left(\muc F \right)^2 }{32 \Dr }. 
\end{equation}
Contrary to the exact short-time diffusivity~\eqref{eq:MSD_smalltime_exact}, the macroscopic result~\eqref{eq:MSD_smalltime_macro} does not depend on the external force. Yet, it is still larger than that of a passive particle ($D$) by the term $\DcM/2$. 
This term also appears in the macroscopic long-time diffusion coefficient 
$\mathcal{D}_{\rm m}$, Eq.~\eqref{eq:D_macro}. Furthermore, $\mathcal{D}_{\rm m}$ depends on the constant part of the macroscopic active velocity $u$ and on the external force amplitude $F$. 

\begingroup 
\def\arraystretch{2.2}%
\begin{table*}
\small
\caption{\label{tab:1} Mobility and short- and long-time diffusivities of the passive Brownian particle (BP, 2nd column), standard active Brownian particle (ABP, 3rd column), and of the current microscopically reversible active Brownian particle (MRABP) model. The 4th column displays corresponding quantities for MRABP in the macroscopic limit~\eqref{eq:ua_macro}, 5th column shows results of microscopic calculations for MRABP.  
} 
  \begin{tabular*}{\textwidth}{@{\extracolsep{\fill}}lllll}
    \hline
   & Passive BP & Standard ABP & Macroscopic MRABP & Microscopic MRABP\\    
    \hline
    Mobility & $\mu$ & $\mu$ & $\mu +\displaystyle \frac{\muc}{2}  $  & $\mu +\displaystyle \frac{1}{2} \frac{\DaO}{\kB T} $ \\[1ex]
\hline	
$\begin{array}{l}\!\!\!\!\! \textrm{Short-time} \\[-2.5ex] \!\!\!\!\! \textrm{diffusion coefficient} \end{array}$
& $D$ & $D$ & $D +\displaystyle \frac{\DcM}{2}$ & $D +\displaystyle \frac{\DaO}{2} \left[1+ \left( \frac{F\delta r}{4\kB T} \right)^2 \right]$ \\
\hline	\\[-4ex]
$\begin{array}{l}\!\!\!\!\! \textrm{Long-time} \\[-2.5ex] \!\!\!\!\! \textrm{diffusion coefficient} \end{array}$
& $D$ & $D +\displaystyle\frac{u^2}{2\Dr} $
& $\begin{array}{l}\!\!\!\!\! D +\displaystyle \frac{\DcM}{2}  + \\ \!\!\!\!\! \phantom{D} + \displaystyle \frac{u^2}{2\Dr} + \frac{\left(\muc F \right)^2 }{32 \Dr } \end{array}$ 
& $\begin{array}{l}\!\!\!\!\! D +\displaystyle \frac{\DaO}{2} \left[1+ \left( \frac{F\delta r}{4\kB T} \right)^2 \right] + \\ \!\!\!\!\! \phantom{D} +\displaystyle  \frac{(\vaO)^2}{2\Dr} \left[1 + \frac{1}{8} \left( \frac{F\delta r}{\kB T} \right)^2 \right] 
 +\frac{1}{32\Dr}  \left( \frac{\DaO F }{\kB T} \right)^2 \end{array}$
\\[6ex]
    \hline
  \end{tabular*}
\end{table*}
\endgroup 

\begin{figure}[b]
\centering
\includegraphics[width=0.85\columnwidth]{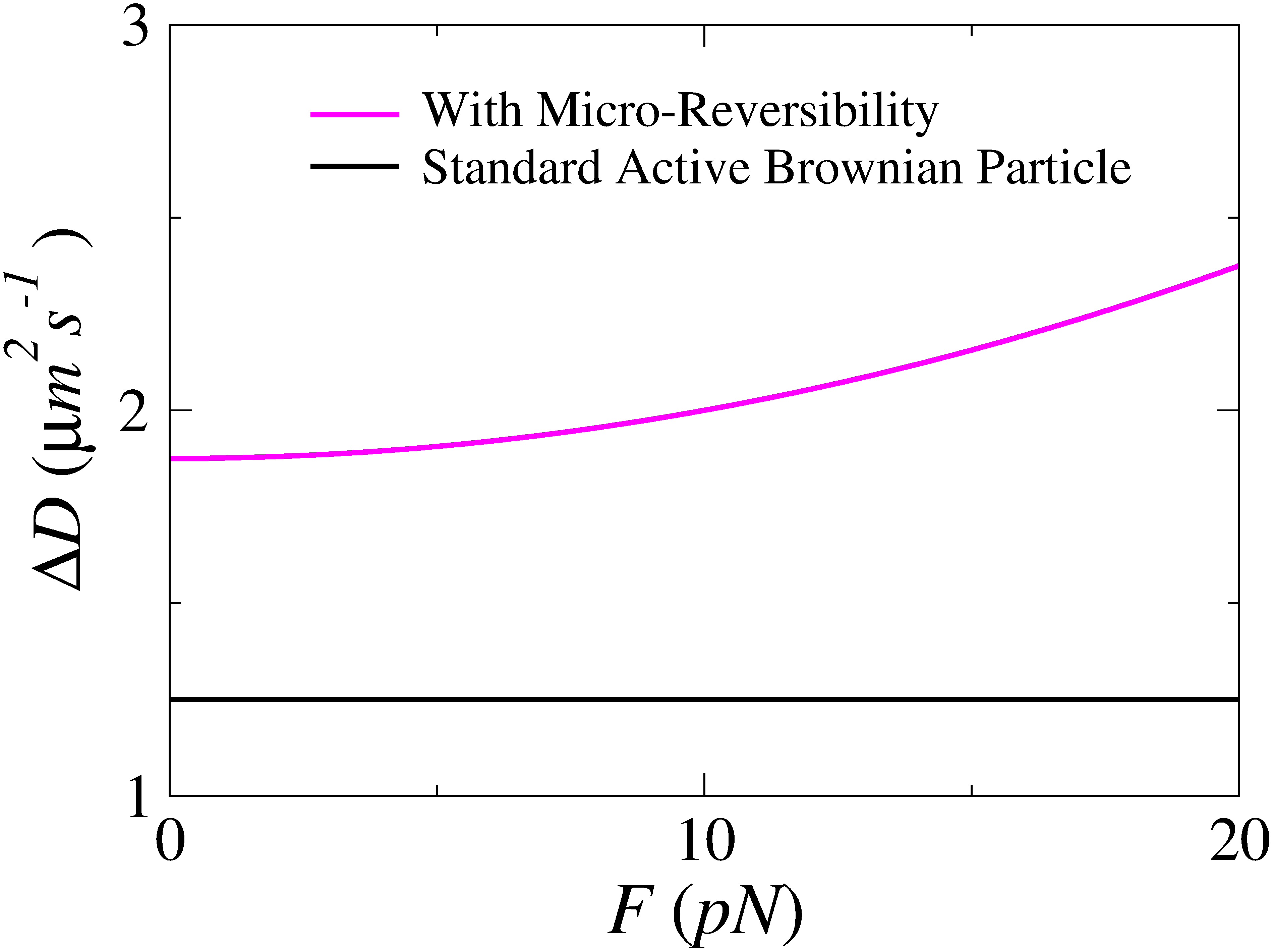}
\caption{An order of magnitude estimation of the increase $\Delta D=\mathcal{D} -D$, with $\mathcal{D} $ given in Eq.~(\ref{eq:D_full}), of the diffusion coefficient of a micro-reversible nanoswimmer as the function of the external force amplitude $F$; magenta curve. Black line shows the increase predicted by the standard active Brownian particle model. In this estimations, we used $\delta r = 5~\mathrm{nm}$, $k_+ =10^5~\textrm{s}^{-1}$, and $\Dr = 10^5~\textrm{s}^{-1}$. }
\label{fig:experimental}
\end{figure}

Finally, we discuss an important special case of our macroscopic results, which is formally obtained by setting $\muc = 0$ in~\eqref{eq:ua_macro}. This is equivalent to neglecting both the MR and the fluctuations of the macroscopic active velocity~\eqref{eq:ua_macro}, which then becomes constant, $u_{\rm a}(t)=u$. 
Langevin equations~\eqref{eq:Langevin_exact} with $u_{\rm a}(t)=u$ describe dynamics of the standard active Brownian particle (ABP) model \cite{SCHIMANSKYGEIER:1995, tenHagen/etal:2011, Romanczuk2012}. In ABP, the mean particle displacement parallel to the force equals to that of a passive particle: 
$\langle x(t) \rangle = \mu F t$. Also the short-time diffusivity for ABP is given by that of a passive particle since $\DcM=0$ in~\eqref{eq:MSD_smalltime_macro}. 
For the long-time diffusion coefficient, the ABP yields 
$\mathcal{D}_{\rm ABP} = D + u^2/2\Dr$, where both $\DcM/2$ and 
$(\muc F)^2/32\Dr$ terms are missing as compared to the thermodynamically consistent macroscopic result~\eqref{eq:D_macro}. 
Consequently, if applied to a particle driven by a microscopically reversible active process, the ABP yields smaller mobility and both diffusivities as compared to the corresponding thermodynamically consistent model. Table~\ref{tab:1} summarizes our predictions for the particle mobility and the diffusivity enhancement (relative to the case of a passive particle) and compares those to the corresponding results of the standard active Brownian particle model. Notice that the standard ABP model has recently been used in analysis of enzyme diffusivities \cite{FENG2019}, where MR and fluctuations of the active velocity can be relevant.

Figure~\ref{fig:experimental} shows an order of magnitude estimate of the predicted diffusion enhancement $\Delta D=\mathcal{D}_{\rm m}-D$ (magenta curve), calculated  according to Eq.~(\ref{eq:D_macro}) as a function of the magnitude $F$ of an external force; black line corresponds to the case of the standard active Brownian model with $\muc = 0$. We use $\delta r = 5~\mathrm{nm}$, $k_+ = 10^5~\textrm{s}^{-1}$, and $\Dr = 10^5~\textrm{s}^{-1}$, which is of the same order as the values reported in experimental studies \cite{Jee:2018a, Ah-Young/etal:2018, Jee:2020} for catalytically active urease.

\section{Summary and implications for experiments}

We have identified several unique features resulting from the microscopic reversibility in the generic model of stochastic nanoswimmer with the microscopically reversible active propulsion. These include enhancements of the mobility and the short- and long-time diffusion constants as compared to the passive dynamics and to the active dynamics not obeying MR. Furthermore, the diffusion constants are force-dependent, which makes our predictions particularly relevant for experiments with nanoswimmers acted upon by external forces. 

In such experiments, measurements of the mean particle position~\eqref{eq:xmean_exact} shall give us the active diffusivity $\DaO$ provided the passive mobility $\mu$ is known.\footnote{The passive mobility can be calculated based on the Stokes relation: $1/\mu = 6 \pi \eta R$, where $\eta$ is the dynamic viscosity and $R$ denotes the hydrodynamic radius of the nanoswimmer.} Alternatively, $\DaO$ can be obtained from the short-time diffusivity~\eqref{eq:MSD_smalltime_exact} at zero external force. The value of $\DaO$ is crucial for understanding the enhancements of the particle diffusion constants in~\eqref{eq:MSD_smalltime_exact} and~\eqref{eq:D_full}. It is present also in the macroscopic approximation, where the discrete nature of active propulsion can be disregarded. At this coarse grained level of description, $\DcM$ (the macroscopic counterpart of $\DaO$) contributes to the enhancement of the overall mobility and to the both diffusivities [Eqs.~\eqref{eq:MSD_smalltime_macro} and \eqref{eq:D_macro}]. If in addition to $\DaO$ the reaction rates $k_\pm^{(0)}$ for the active propulsion are known, we can estimate the elementary active displacement $\delta r$ based on the definition of $\DaO$ in  \eqref{eq:DaO}. This estimate can give us the mean active velocity $\vaO$~\eqref{eq:vaO}, which also contributes to the enhancement of the long-time diffusion constant~\eqref{eq:D_full}. 

\section{Conclusions}

From a general perspective, the demonstrated sensitivity of diffusion coefficients and the mobility to the coupling between chemical and mechanical degrees of freedom can contribute to the basic understanding of diffusivity enhancement of individual active molecules. Indeed, those operate on time scales where fluctuations in chemical reactions and their statistical properties can be relevant, therefore correctly including the microscopic reversibility can be crucial in this case, in contrast to micron-sized colloids powered by tenths of thousands of reaction steps per an elementary displacement. 

The force-dependence of diffusivities indicates a way to experimentally verify the role of MR by varying the amplitude of the externally applied force. Moreover, according to these results, the diffusion constants can be sensitive also to local forces. For example, in experiments where nanoswimmers move near surfaces, electrostatic or van der Waals interactions with the surface can have non-negligible effects on the measured diffusion constants. Performing such experiments in a controlled manner can yield a definite answer regarding relevance of MR for the active propulsion at the nanoscale, which in turn can contribute to better understanding of further physical mechanisms determining the value of particle's diffusivity.

\section*{Acknowledgements}
We acknowledge financial support from the Portuguese Foundation for Science and Technology (FCT) under Contracts nos.\ PTDC/FIS-MAC/5689/2020, PTDC/FIS-MAC/28146/2017 (LISBOA-01-0145-FEDER-028146), UIDB/00618/2020, UIDP/00618/2020, and IF/00322/2015. AR gratefully acknowledges financial support by the Czech Science Foundation (Project No.\ 20-02955J).

%
\end{document}